\documentclass[twocolumn,showpacs]{revtex4}

\usepackage{graphicx}
\usepackage{dcolumn}
\usepackage{amsmath}

\begin{document}

\title{Phase diagram of the $t$-$U^2$ Hamiltonian of the weak coupling
Hubbard model
}

\author{Takashi Yanagisawa$^{a,b}$}

\affiliation{$^a$Condensed-Matter Physics Group, Nanoelectronics Research
Institute,
National Institute of Advanced Industrial Science and Technology (AIST)
Central 2, 1-1-1 Umezono, Tsukuba 305-8568, Japan\\
$^b$CREST, Japan Science and Technology Agency (JST), Kawaguchi,
Saitama 332-0012, Japan}

\date{}

\begin{abstract}
We determine the symmetry of Cooper pairs, on the basis of the perturbation
theory in terms of the Coulomb interaction $U$, for the two-dimensional Hubbard 
model on the square lattice.
The phase diagram is investigated in detail.
The Hubbard model for small $U$ is mapped onto an effective Hamiltonian with the attractive
interaction
using the canonical transformation: $H_{eff}=e^S He^{-S}$.
The gap equation of the weak coupling formulation is solved without numerical ambiguity 
to determine the
symmetry of Cooper pairs.
The superconducting gap crucially depends on the position of the van Hove singularity.
We show the phase diagram in the plane of the electron filling $n_e$ and the next
nearest-neighbor transfer $t'$.
The $d$-wave pairing is dominant for the square lattice in a wide range of
$n_e$ and $t'$.
The $d$-wave pairing is also stable for the square lattice with anisotropic $t'$.
The three-band $d$-$p$ model is also investigated, for which the $d$-wave
pairing is stable in a wide range of $n_e$ and $t_{pp}$ (the transfer between
neighboring oxygen atoms).
In the weak coupling analysis, the second-neighbor transfer parameter $-t'$
could not be so large so that
the optimum doping rate is in the range of $0.8 <n_e <0.85$.
\end{abstract}


\maketitle

\section{Introduction}

Since the discovery of high-temperature superconductors, the strongly correlated
electron systems have been studied intensively.
The effect of the strong correlation between electrons is important for many
quantum critical phenomena such as unconventional superconductivity (SC).
High-temperature
superconductors\cite{dag94,sca90,and97} as well as
heavy fermions\cite{ste84,lee86,ott87,map00} are known as the typical 
correlated electron systems.
These systems are modeled by the Hamiltonian with the electronic interaction
of the on-site Coulomb repulsion.
Recently the mechanisms of
superconductivity in high-temperature superconductors
have been extensively studied using the two-dimensional Hubbard
model\cite{loh90,mor91,whe93,nak97,yam98,neu03,yan01,yan02,miy04}.

The superconductivity of the Hubbard model has been questioned for many
years.
It is extremely difficult to show the existence of superconducting
phase for the Hubbard model in a reasonable way.
For the present we cannot answer this long-standing question soon.
Instead of examining the possibility of superconductivity,
it is possible to investigate possible symmetries of Cooper pairs for
an effective Hamiltonian with the attractive interaction.
For this purpose
effective Hamiltonians have been obtained for the Hubbard model.
The t-J model is the well known effective Hamiltonian derived in the
limit of the large on-site repulsion $U$, using the canonical
transformation $H_{t-J}=e^S He^{-S}$ with $S\propto t/U$.
On the other hand, in the limit of small $U$, the perturbation theory
also leads to an effective Hamiltonian with the attractive
interaction\cite{miy86,sca86,kon01,hlu99}, where we have $S\propto U/t$.
The phase diagram with respect to the Cooper pair symmetry can be determined
if we solve the gap equation.

We must notice that we should compare the energy
with other electronic states to show that the superconducting state
is indeed stable.
For the half-filled band with vanishing $t'=0$ in two space dimensions, the 
antiferromagnetic order parameter for small $U$ is\cite{sch02}
\begin{equation}
\Delta_{AF}=\frac{8tc}{U}{\rm exp}\left( -\sqrt{\frac{4\pi^2tc}{U}} \right),
\end{equation}
where $c=3-\sqrt{3}$.
It is, however, obvious that the antiferromagnetically ordered state is unstable 
away from half filling if the
Coulomb repulsion $U$ is small.
Thus we focus on the case of small $U$ for which we have also a merit that
the gap equation is considerably simplified.
The purpose of the paper is to determine the gap symmetry for the 
square lattice using the
small-$U$ gap equation derived for the effective Hamiltonian.
Although the real superconductivity in correlated electron systems should
be described by a theory of strong-coupling superconductivity, 
the phase diagram can be determined in detail using the weak coupling formulation.
Precise calculations are sometimes not easy at low temperatures
in the strong-coupling formulation
due to the Matsubara frequency summation and the wave number summation.
It is important to examine the detailed phase diagram for materials belonging to
strongly correlated systems such as the cuprate high temperature superconductors,
the organic superconductors, the ruthenate superconductor Sr$_2$RuO$_4$.

The paper is organized as follows.
In Section II the effective Hamiltonian is derived using the canonical
transformation.  We show that we can derive the attractive effective
Hamiltonian using some approximations. 
In Section III the gap equation is shown and the results are presented
in Section IV.
We give a summary in Section V.

\section{Effective Hamiltonian}

The Hubbard Hamiltonian is
\begin{eqnarray}
H&=& -t\sum_{\langle ij\rangle\sigma}(c_{i\sigma}^{\dag}c_{j\sigma}
+{\rm h.c.})
-t'\sum_{\ll j\ell\gg\sigma}(c_{j\sigma}^{\dag}c_{\ell\sigma}
+ {\rm h.c.})\nonumber\\
&+& U\sum_i n_{i\uparrow}n_{i\downarrow}
\end{eqnarray}
where $\langle ij\rangle$ and $\ll j\ell\gg$ denote the nearest-neighbor 
and next-nearest-neighbor pairs, respectively.
$U$ is the on-site Coulomb repulsion.
The unit of energy is given by $t$ in this paper.
The total number of sites and the number of electrons are denoted as
$N$ and $N_e$, respectively.
The half-filled band corresponds to $n_e=N_e/N=1$.

The effective Hamiltonian is derived using the perturbation theory for
small $U$.
The canonical transformation also maps the Hubbard
model to an effective Hamiltonian with the attractive interaction\cite{konp}.
Since no instability except superconductivity occurs for small $U$ away from
half filling, we assume that the pairing interaction is the most singular.
The procedure of mapping is as follows.
The Hamiltonian is written as
\begin{equation}
H= H_0+H_1+H_2+H_3,
\end{equation}
where
\begin{equation}
H_0= \sum_{k\sigma}\xi_k c^{\dag}_{k\sigma}c_{k\sigma},
\end{equation}
\begin{equation}
H_1= \frac{U}{N}\sum_{k\ne k'} c^{\dag}_{k'\uparrow}c^{\dag}_{-k'\downarrow}
c_{-k\downarrow}c_{k\uparrow},
\end{equation}
\begin{equation}
H_2= \frac{U}{N}\sum_{k\ne k',q\ne 0} c^{\dag}_{k'\uparrow}
c^{\dag}_{-k'-q\downarrow}
c_{-k-q\downarrow}c_{k\uparrow},
\end{equation}
\begin{equation}
H_3= \frac{U}{N}\sum_{kk'} c^{\dag}_{k\uparrow}c_{k\uparrow}
c^{\dag}_{k'\downarrow}c_{k'\downarrow}.
\end{equation}
The dispersion relation  $\epsilon_k$ for the square lattice is
\begin{equation}
\epsilon_k= -2t({\rm cos}k_x+{\rm cos}k_y)
-4t'{\rm cos}k_x{\rm cos}k_y,
\end{equation}
where $\mu$ is the chemical potential.
We set $\xi_k=\epsilon_k-\mu$.
Using a canonical transformation, $\tilde{\psi}=e^S \psi$, we look for
the solution of the Schr\"{o}dinger equation 
$H_{eff}\tilde{\psi}=E\tilde{\psi}$.
The effective Hamiltonian reads
\begin{eqnarray}
H_{eff}&=& e^S He^{-S}=H+[S,H]+\frac{1}{2}[S,[S,H]]+\cdots\nonumber\\
&=& H_0+H_1+H_2+H_3+[S,H_0+H_1+H_2+H_3]\nonumber\\
&+&\frac{1}{2}[S,[S,H_0]]+\cdots.
\end{eqnarray}
We determine $S$ so as to satisfy $H_2+[S,H_0]=0$.  We find
\begin{equation}
S= \frac{U}{N}\sum_{k\neq k',q\neq 0}\frac{1}{\xi_{k'+q}+\xi_{k'}
-\xi_{k+q}-\xi_k}
\cdot c^{\dag}_{k'\uparrow}c_{-k'-q\downarrow}
c^{\dag}_{-k-q\downarrow}c_{k\uparrow}.
\end{equation}
Since $[S,H_3]=0$, we obtain up to the order of $U^2$,
\begin{equation}
H_{eff}= H_0+H_1+H_3+[S,H_1]+\frac{1}{2}[S,H_2].
\end{equation}
The commutator $[S,H_2]$ is evaluated as
\begin{eqnarray}
[S,H_2]&=& \left(\frac{U}{N}\right)^2
\sum_{k\neq k',q\neq 0}\sum_{p\neq p',q'\neq 0}S_{kk'}^q\nonumber\\
&\times&(-\delta_{pk'}c_{p'\uparrow}^{\dag}c_{-p'-q'\downarrow}^{\dag}
c_{-p-q'\downarrow}c_{-k'-q\downarrow}^{\dag}c_{-k-q\downarrow}c_{k\uparrow}\nonumber\\
&&+\delta_{p+q',k'+q}c_{p'\uparrow}^{\dag}c_{-p'-q'\downarrow}^{\dag}c_{k'\uparrow}^{\dag}
c_{p\uparrow}c_{-k-q\downarrow}c_{k\uparrow}\nonumber\\
&&-\delta_{p'+q',k+q}c_{k'\uparrow}^{\dag}c_{-k'-q\downarrow}^{\dag}c_{p'\uparrow}^{\dag}
c_{k\uparrow}c_{-p-q'\downarrow}c_{p\uparrow}\nonumber\\
&&+\delta_{p'k}c_{k'\uparrow}^{\dag}c_{-k'-q\downarrow}^{\dag}c_{-k-q\downarrow}
c_{-p'-q'\downarrow}^{\dag}c_{-p-q'\downarrow}c_{p\uparrow}),\nonumber\\
\end{eqnarray}
where
\begin{equation}
S_{kk'}^q = \frac{1}{\xi_{k'+q}+\xi_{k'}-\xi_{k+q}-\xi_k}.
\end{equation}
Since the purpose of this paper is to investigate the pairing symmetry,
we need only the first term and the last term of $[S,H_2]$.
We find that the average of the second and third terms with respect to the BCS wave
function vanish.  Due to the same reason $[S,H_1]$ can be neglected.
Then the effective Hamiltonian is
\begin{eqnarray}
H_{t-U^2}&=& \sum_{k\sigma}\xi_k c^{\dag}_{k\sigma}c_{k\sigma}
+\frac{U}{N}\sum_{k\neq k'}c^{\dag}_{k'\uparrow}c^{\dag}_{-k'\downarrow}
c_{-k\downarrow}c_{k\uparrow}\nonumber\\
&+&\frac{1}{2}\frac{U^2}{N^2}\sum_{k\neq k',q\neq 0}\sum_{p\neq p',q'\neq 0}
\frac{1}{\xi_{k'+q}+\xi_{k'}-\xi_{k+q}-\xi_k}\nonumber\\
&\cdot&(\delta_{p'k}c^{\dag}_{k'\uparrow}c_{-k-q\downarrow}c^{\dag}_{-k'-q\downarrow}
c_{-p-q'\downarrow}c^{\dag}_{-p'-q'\downarrow}c_{p\uparrow}\nonumber\\
&-&\delta_{pk'}c^{\dag}_{p'\uparrow}c_{-p-q'\downarrow}c^{\dag}_{-p'-q'\downarrow}
c_{-k-q\downarrow}c^{\dag}_{-k'-q\downarrow}c_{k\uparrow} ),\nonumber\\
\end{eqnarray}

If we set $k=p+q'$ and $p'=k'+q$, the first term of $[S,H_2]$ is approximated as
\begin{eqnarray}
H_{2a}&\equiv& -\left(\frac{U}{N}\right)^2\sum_{k\neq k',q\neq 0}
\sum_{p\neq p',q'\neq 0}S_{k'k}^q \delta_{pk'}\nonumber\\
&\times& c_{p'\uparrow}^{\dag}c_{-p'-q'\downarrow}^{\dag}
c_{-p-q'\downarrow}c_{-k'-q\downarrow}^{\dag}c_{-k-q\downarrow}c_{k\uparrow}\nonumber\\
&\approx& \left(\frac{U}{N}\right)^2\sum_{k\neq k',q\neq 0}
c^{\dag}_{k'+q\uparrow}c^{\dag}_{-k'-q\downarrow}
 \frac{c^{\dag}_{-k-q\downarrow}c_{-k-q\downarrow}}
{\xi_{k'+q}+\xi_{k'}-\xi_{k+q}-\xi_k}\nonumber\\
&\times&c_{-k\downarrow}c_{k\uparrow}\nonumber\\ 
&\approx& \left(\frac{U}{N}\right)^2\sum_{k\neq k'-q,q\neq 0}
c^{\dag}_{k'\uparrow}c^{\dag}_{-k'\downarrow}
 \frac{f_{-k-q}}
{\xi_{k'}+\xi_{k'-q}-\xi_{k+q}-\xi_k}\nonumber\\
&\times& c_{-k\downarrow}c_{k\uparrow}
\nonumber\\
&=& \left(\frac{U}{N}\right)^2\sum_{k+q\neq 0,k'+q\neq 0}
c^{\dag}_{k'\uparrow}c^{\dag}_{-k'\downarrow}\nonumber\\
&\times& \frac{f_{q}}
{\xi_{k'+k+q}-\xi_{-q}+\xi_{k'}-\xi_k}
c_{-k\downarrow}c_{k\uparrow}
\end{eqnarray}
where $f_k$ is the Fermi distribution function,
\begin{equation}
f_k= \frac{1}{ {\rm e}^{\beta\xi_k}+1 }.
\end{equation}
Since the summation is restricted to the small region near the
Fermi surface, we obtain assuming $\xi_{-k}=\xi_k$
\begin{equation}
H_{2a}\approx \left(\frac{U}{N}\right)^2\sum_{k\neq -q,k'\neq -q}
\frac{f_q}{\xi_{k'+k+q}-\xi_q}
 c^{\dag}_{k'\uparrow}
c^{\dag}_{-k'\downarrow}c_{-k\downarrow}c_{k\uparrow}.
\end{equation}
Similarly the last term of $[S,H_2]$ is written as
\begin{equation}
H_{2b}\approx \left(\frac{U}{N}\right)^2\sum_{k\neq -q,k'\neq -q}
\frac{f_{k'+k+q}}{\xi_{k'+k+q}-\xi_q}
 c^{\dag}_{k'\uparrow}c^{\dag}_{-k'\downarrow}c_{-k\downarrow}
c_{k\uparrow}.
\end{equation}
The resulting effective Hamiltonian is
\begin{eqnarray}
H_{eff}&=& e^S He^{-S}\equiv H_{t-U^2}\nonumber\\
&=& \sum_{k\sigma}\xi_kc^{\dag}_{k\sigma}c_{k\sigma}
+\sum_{kk'}V_{kk'}c^{\dag}_{k'\uparrow}c^{\dag}_{-k'\downarrow}
c_{-k\downarrow}c_{k\uparrow},\nonumber\\
\end{eqnarray}
where
\begin{equation}
V_{kk'}=\frac{U}{N}+\frac{U^2}{N}\chi({\bf k}+{\bf k'}).
\end{equation}
$\chi(k+k')$ is the magnetic susceptibility defined as
\begin{equation}
\chi({\bf k}+{\bf k'})= \frac{1}{N}\sum_q
\frac{f_{k+k'+q}-f_q}{\xi_q-\xi_{k+k'+q}}.
\end{equation}
Thus we have reached the effective Hamiltonian up to the order of
$U^2$ using the canonical transformation.

\section{Gap equation}
The gap equation for the $t$-$U^2$ model was investigated in Ref.\cite{kon01}.
Since the equation was considerably simplified for small $U$, the gap
equation was solved without numerical ambiguity.
We define the order parameter,
\begin{equation}
\Delta_k= \sum_{k'}V_{kk'}\langle c_{-k'\downarrow}c_{k'\uparrow}\rangle.
\end{equation}
Using the mean-field theory, the gap equation for the Hamiltonian 
$H_{t-U^2}$ is
\begin{equation}
\Delta_k= -\sum_{k'}V_{kk'}\Delta_{k'}\frac{1-2f(E_{k'})}{2E_{k'}},
\end{equation}
where $E_k=\sqrt{\xi_k^2+\Delta_k^2}$.
We assume the anisotropic order parameter given as
\begin{equation}
\Delta_{{\bf k}}=\Delta\cdot z_{{\bf k}},
\end{equation}
where $z_{{\bf k}}$ denotes the ${\bf k}$-dependence of $\Delta_{{\bf k}}$.
At $T=0$ the gap equation is written as
\begin{equation}
\Delta_k=-\frac{1}{2}\sum_{k'}V_{kk'}\frac{\Delta_{k'}}{E_{k'}}.
\end{equation}
For small $U$, the gap equation for anisotropic pairing is extremely 
simplified retaining only the logarithmic term\cite{kon01}:
\begin{equation}
z_{k}= {\rm log}\left(\frac{\Delta}{2\omega_0}\right)U^2\frac{1}{N}
\sum_{k'}\chi({\bf k}+{\bf k}')\delta(\xi_{k'})z_{k'},
\end{equation}
where $\omega_0$ is the cut-off energy.
The critical temperature $T_c$ is determined by
\begin{equation}
z_{{\bf k}}=-\sum_{{\bf k}'}V_{{\bf k}{\bf k}'}z_{{\bf k}'}
\frac{1-2f(\left|\xi_{{\bf k}'}\right|)}{2\left|\xi_{{\bf k}'}\right|},
\end{equation}
for $T=T_c$.
For small $U$, the critical temperature $T_c$ is extremely small.
In this case we can use the following approximation,
\begin{eqnarray}
I&\equiv& \int_0^{\omega_0}d\xi g(\xi)\frac{{\rm tanh}(\beta_c\xi/2)}{\xi}\nonumber\\
&\approx& g(\omega_0){\rm log}\omega_0
-\frac{\beta_c}{2}\int_0^{\omega_0}d\xi{\rm log}\xi\cdot g(\xi)
\frac{1}{({\cosh}(\beta_c\xi/2))^2}\nonumber\\
&=& g(\omega_0){\rm log}\omega_0
-\int_0^{\beta_c\omega_0/2}dx{\log}\left(\frac{2x}{\beta_c}\right)
g\left(\frac{2x}{\beta_c}\right)\frac{1}{({\rm cosh}x)^2}\nonumber\\
&=& g(\omega_0){\rm log}\omega_0
-g(0)\int_0^{\beta_c\omega_0/2}dx{\log}\left(\frac{2x}{\beta_c}\right)
\frac{1}{({\rm cosh}x)^2}\nonumber\\
&\approx& g(0){\rm log}\left(\frac{2{\rm e}^{\gamma}\omega_0}
{\pi k_BT_c}\right), 
\end{eqnarray}
where we assume that $g(\xi)$ is a slowly varying function and $g'(\xi)$ 
is negligible.
The equation is written as
\begin{equation}
z_{{\bf k}}=-{\rm log}\left(\frac{2{\rm e}^{\gamma}\hbar\omega_0}{\pi k_BT_c}
\right)\sum_{{\bf k}'}V_{{\bf k}{\bf k}'}z_{{\bf k}'}\delta(\xi_{{\bf k}'}).
\end{equation}
Since $T_c$ is very small, the summation over ${\bf k}'$ can be restricted to the
average over near the Fermi surface.
If we solve the eigenequation
\begin{equation}
\frac{2}{N}\sum_{{\bf k}'}\chi({\bf k}+{\bf k}')z_{{\bf k}'}\delta(\xi_{{\bf k}'})
=-xz_{{\bf k}},
\end{equation}
the critical temperature is obtained as
\begin{equation}
k_BT_c= 1.13\omega_0{\rm exp}\left(-\frac{2t^2}{xU^2}\right),
\end{equation}
where the energy unit is given by $t$.
Since $\Delta$ is given as
\begin{equation}
\Delta= 2\omega_0{\rm exp}\left(-\frac{2t^2}{xU^2}\right),
\end{equation}
the ratio $2\Delta/k_BT_c$ equals the BCS universal value
$2\pi/{\rm e}^{\gamma}=3.53$.

\section{Pairing symmetry}
\subsection{Method of solving the eigenvalue equation}
We express $z_k$ and $V_{kk'}$ in terms of the polar 
coordinates\cite{kon01}:
\begin{equation}
z_k= z(\xi,\theta),
\end{equation}
\begin{equation}
\chi({\bf k}+{\bf k}')=\chi(\xi,\theta,\xi',\theta'),
\end{equation}
where ${\bf k}$ is expressed using the polar angle $\theta$: 
${\bf k}=(\xi,\theta)$ in terms of the polar coordinates. 
We consider the gap function on the Fermi surface 
$z(\theta)\equiv z(0,\theta)$.
If we define $\chi(\theta,\theta')=\chi(0,\theta,0,\theta')$, the gap
equation is
\begin{equation}
2\int_0^{2\pi} d\theta'\rho_F(\theta')\chi(\theta,\theta')z(\theta')=
-xz(\theta),
\end{equation}
where $\rho_F(\theta)$ is the density of states at the Fermi surface:
\begin{equation}
\rho_F(\theta)=\frac{1}{(2\pi)^2}k_F(\theta)
\frac{1}{|\frac{\partial\xi}{\partial k}(k=k_F(\theta))|},
\end{equation}
where $k_F(\theta)$ is the Fermi wave number of the polar coordinate
$\theta$ and $\partial\xi/\partial k$ is the derivative
with respect to $k=|{\bf k}|$.
If we expand $z(\theta)$ as
\begin{equation}
z(\theta)= \sum_n z_n e^{in\theta},
\end{equation}
the gap equation is given as
\begin{equation}
\sum_n \chi_{mn}z_n= -xz_m,
\end{equation}
where $\chi_{mn}$ are the matrix elements of $\chi(\theta,\theta')$:
\begin{equation}
\chi_{mn}= \frac{1}{\pi}\int_0^{2\pi}d\theta d\theta'
\rho_F(\theta')e^{-im\theta}\chi(\theta,\theta')e^{in\theta'}.
\end{equation}
The number of basis functions kept in solving the eigenequation is
30 to 40 in this paper.
The ${\bf k}$-space is divided into $200\times 20$ points on equally
spaced mesh in the
numerical calculations of $\chi(\theta,\theta')$.

\begin{figure}
\begin{center}
\includegraphics[width=\columnwidth]{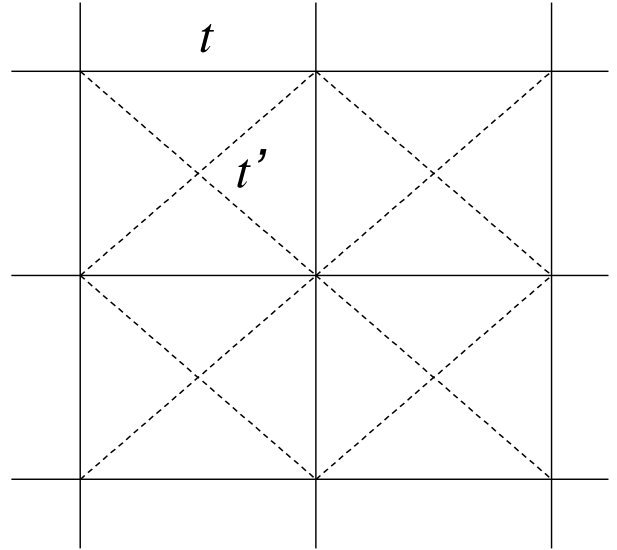}
\caption{
Square lattice with next-nearest transfer $t'$.
}
\end{center}
\label{latt-sq}
\end{figure}

\begin{figure}
\begin{center}
\includegraphics[width=\columnwidth]{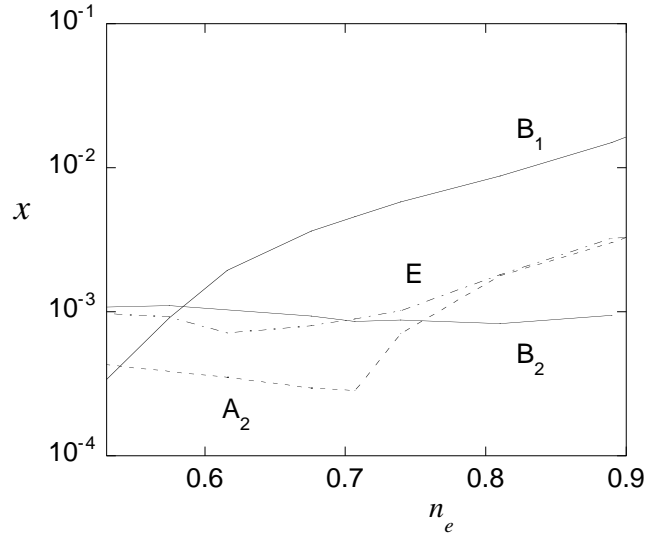}
\caption{
The exponent $x$ as a function of the electron density for
$t'=0$ . (See \cite{kon01}.  We have included $x$ for the E representation.)  
Since the line for A$_1$ mostly coincides with that for B$_2$,
the A$_1$ line is omitted.
}
\end{center}
\label{xne}
\end{figure}

\begin{figure}
\begin{center}
\includegraphics[width=\columnwidth]{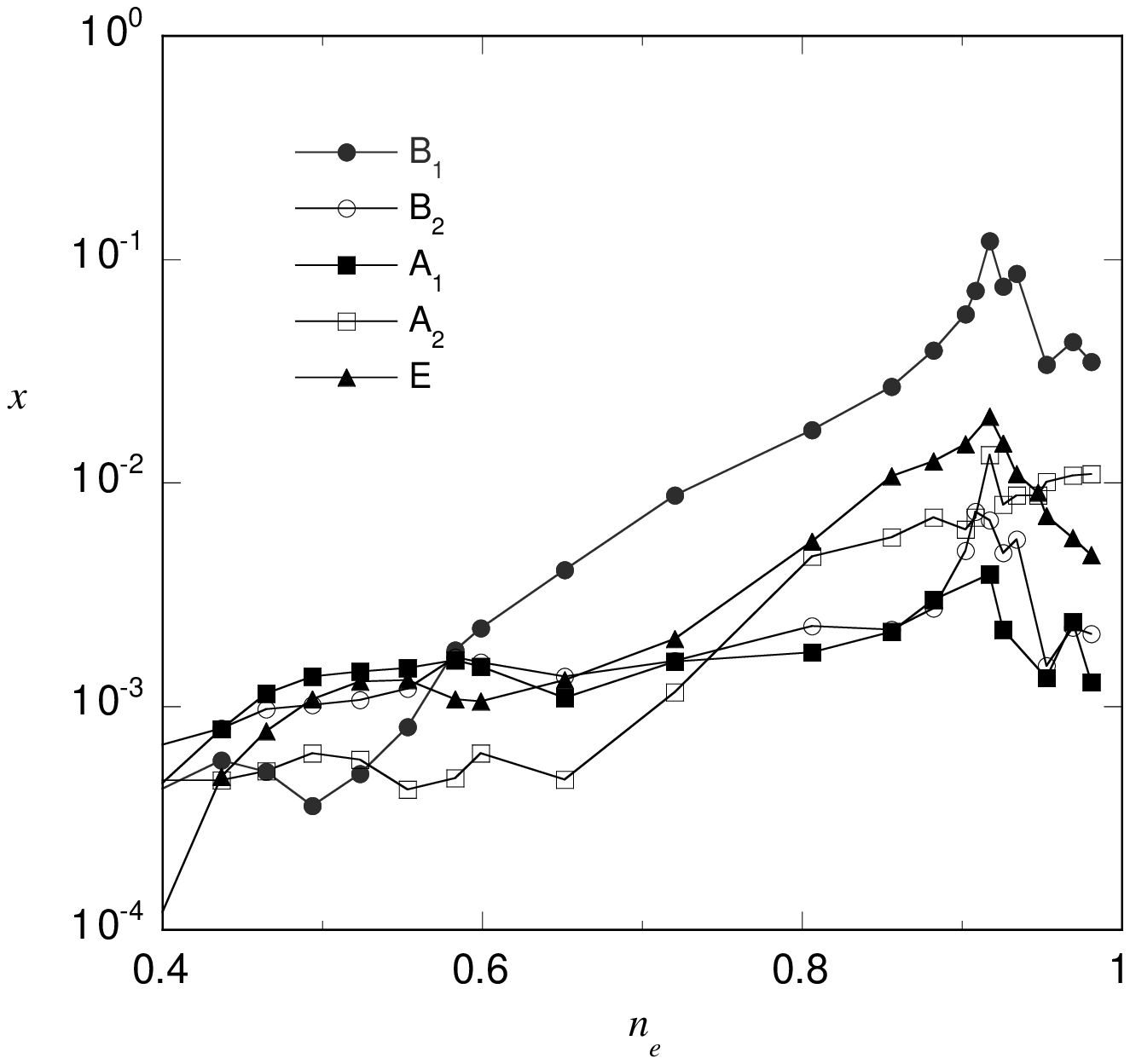}
\caption{
The exponent $x$ as a function of the electron density for
$t'=-0.1$.
}
\end{center}
\label{xne01}
\end{figure}

\begin{figure}
\begin{center}
\includegraphics[width=\columnwidth]{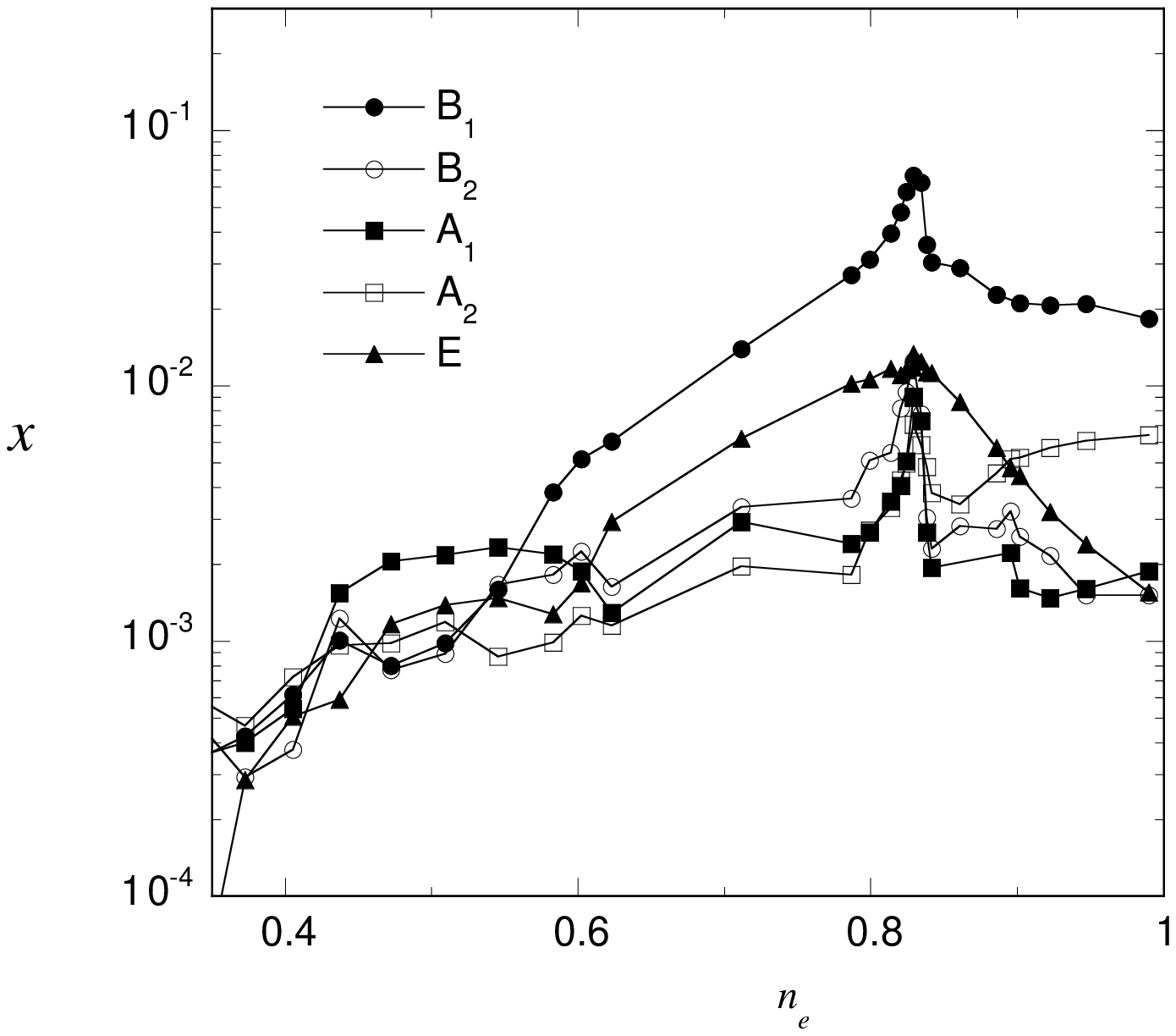}
\caption{
The exponent $x$ as a function of the electron density for
$t'=-0.2$.
}
\end{center}
\label{xne02}
\end{figure}

\begin{figure}
\begin{center}
\includegraphics[width=\columnwidth]{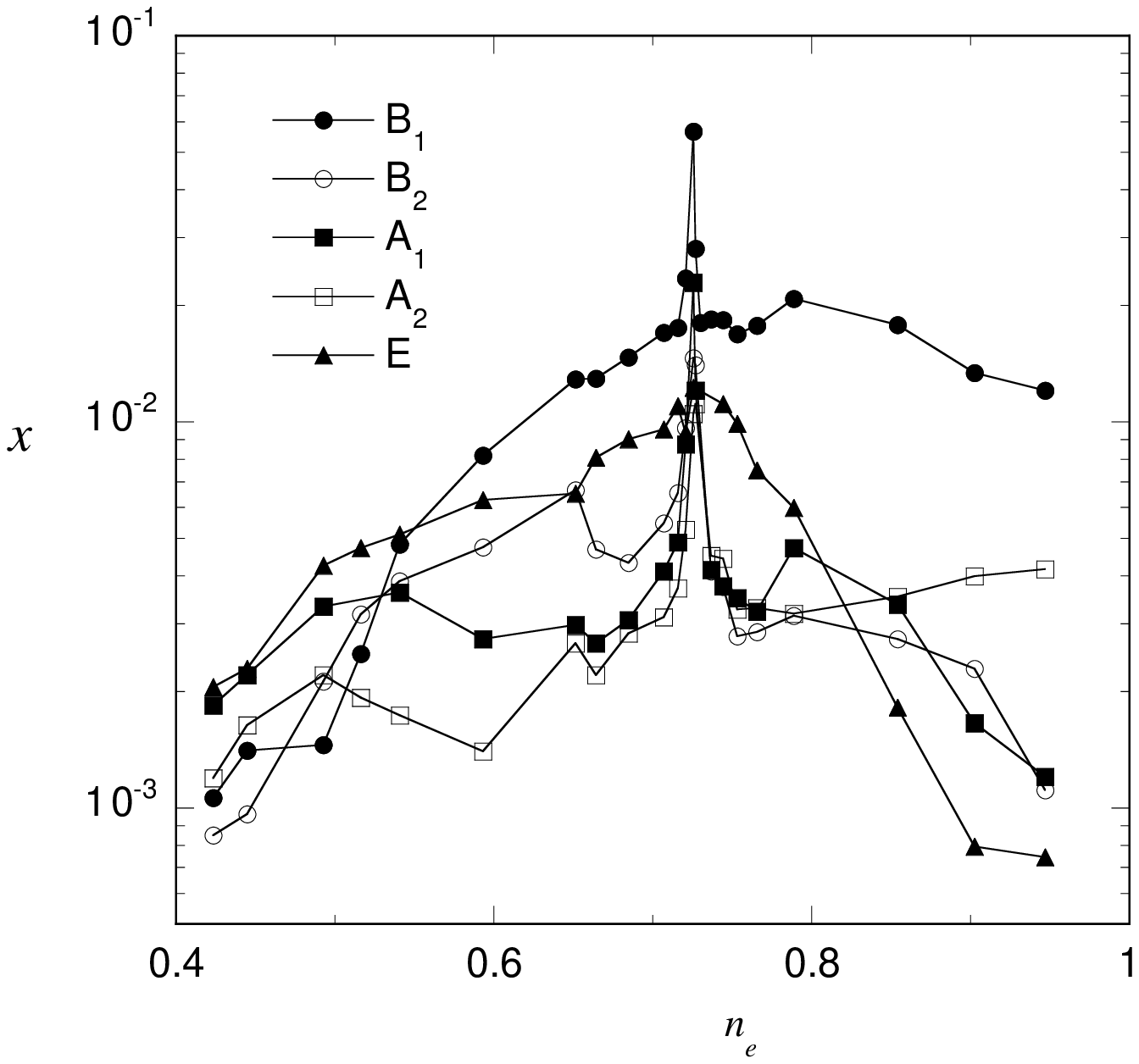}
\caption{
The exponent $x$ as a function of the electron density for
$t'=-0.3$.
}
\end{center}
\label{xne03}
\end{figure}

\begin{figure}
\begin{center}
\includegraphics[width=\columnwidth]{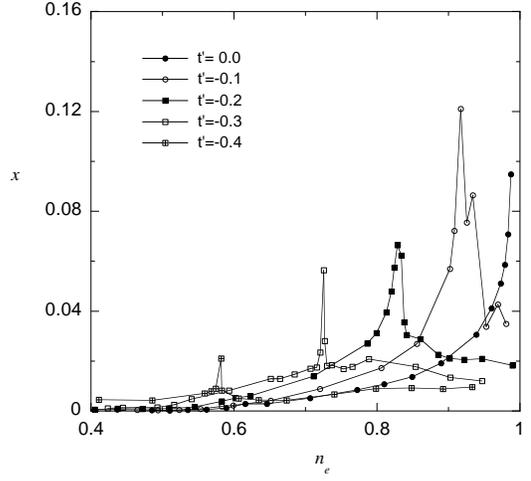}
\caption{
The exponent $x$ of B$_1$ symmetry as a function of the electron density for
$t'=0$, $-0.1$, $-0.2$, $-0.3$ and $-0.4$.
}
\end{center}
\label{xnet'}
\end{figure}

\begin{figure}
\begin{center}
\includegraphics[width=\columnwidth]{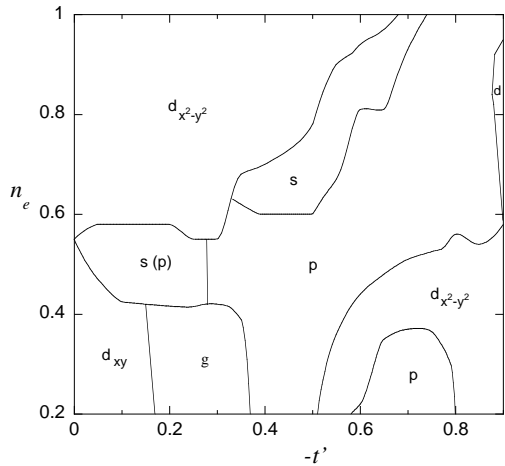}
\caption{
Phase diagram in the $n_e$-$t'$ plane for $t'\leq 0$.
$s$ denotes the pairing state with extended-$s$ wave symmetry.
In the $s$-wave region for small $|t'|$, the $s$- and $p$-wave states are
sometimes nearly degenerate.
Small regions near boundaries are not shown.
}
\end{center}
\label{phase1}
\end{figure}

\begin{figure}
\begin{center}
\includegraphics[width=\columnwidth]{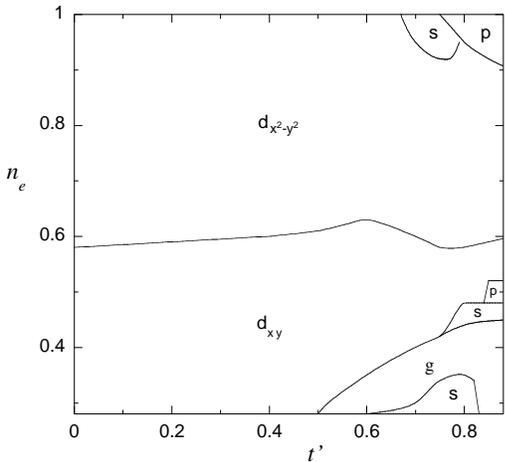}
\caption{
Phase diagram in the $n_e$-$t'$ plane for $t'\geq 0$.
$s$, $g$ and $d_{x^2-y^2}$ wave pairing states are almost degenerate in 
the low-carrier
region for large $t'$.
}
\end{center}
\label{phase2}
\end{figure}

\begin{figure}
\begin{center}
\includegraphics[width=\columnwidth]{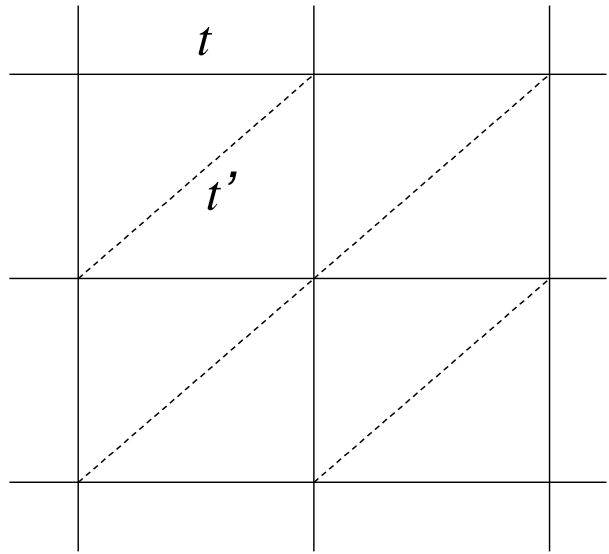}
\caption{
Square lattice with anisotropic next-nearest transfer $t'$
(anisotropic triangular lattice) which is the lattice of organic
conductors.
}
\end{center}
\label{latt-org}
\end{figure}

\begin{figure}
\begin{center}
\includegraphics[width=\columnwidth]{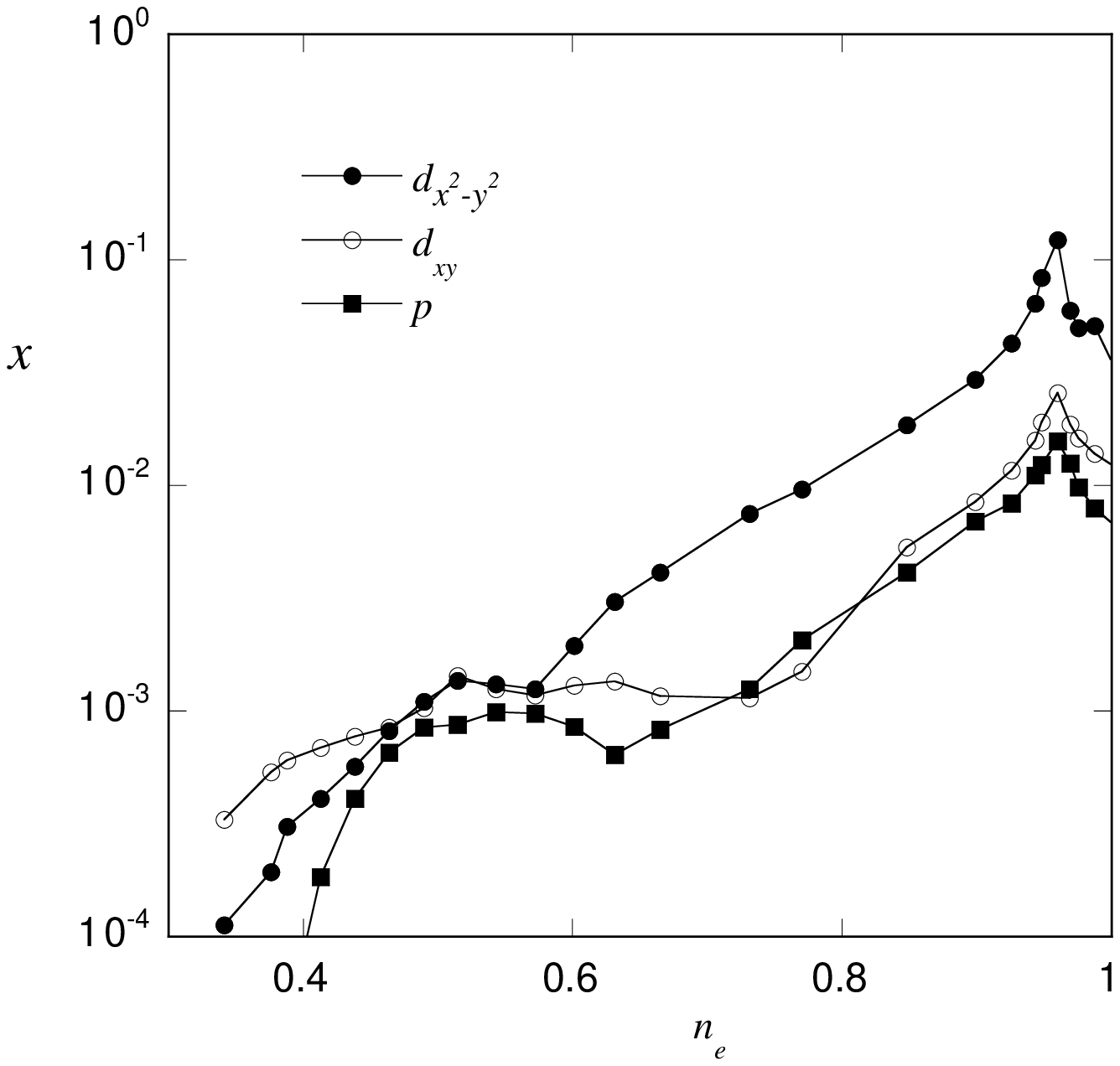}
\caption{
The exponent $x$ on the square lattice with anisotropic $t'=-0.1$.
}
\end{center}
\label{xneor01}
\end{figure}

\begin{figure}
\begin{center}
\includegraphics[width=\columnwidth]{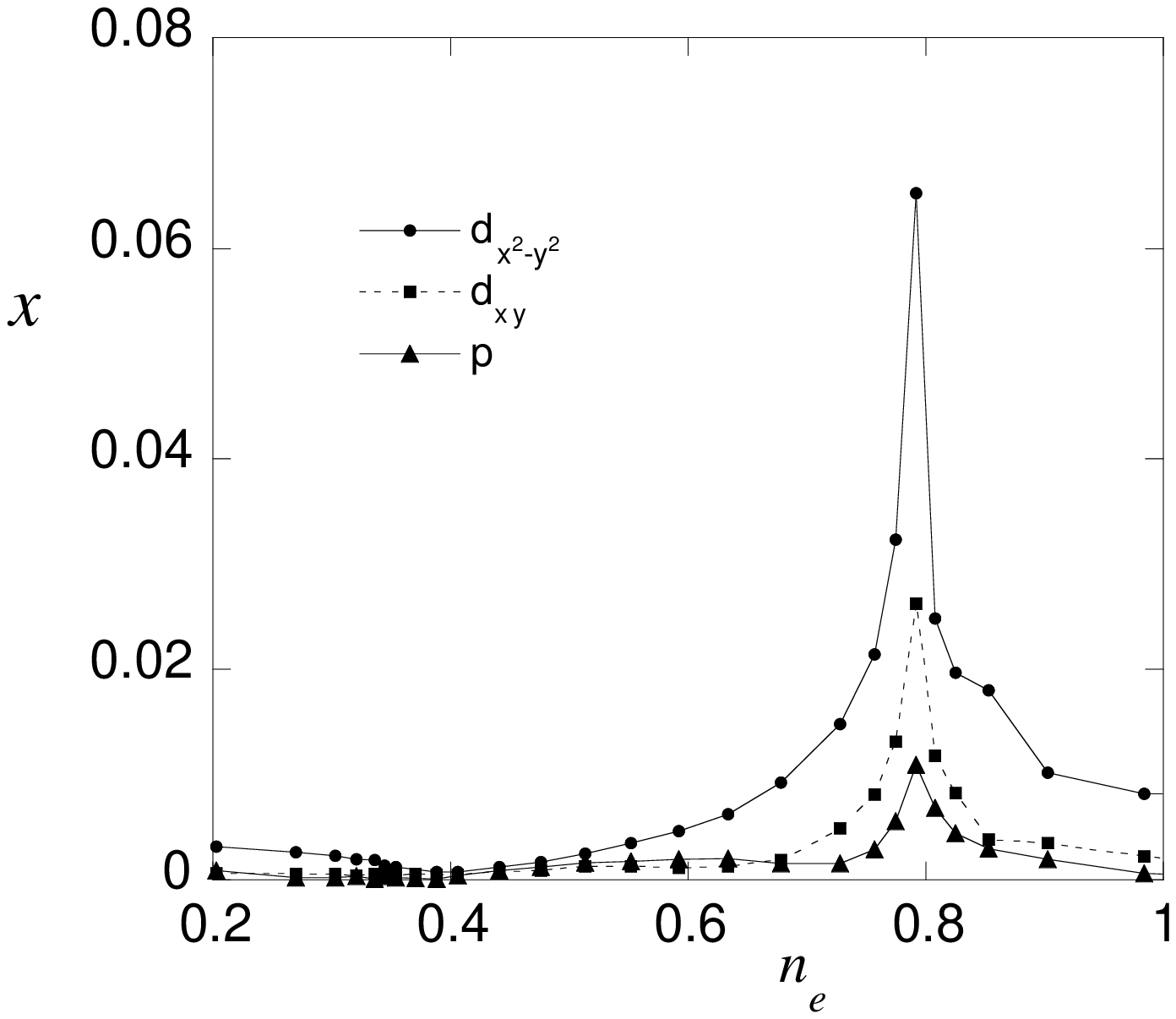}
\caption{
The exponent $x$ on the square lattice with anisotropic $t'=-0.5$.
}
\end{center}
\label{xneor05}
\end{figure}

\begin{figure}
\begin{center}
\includegraphics[width=\columnwidth]{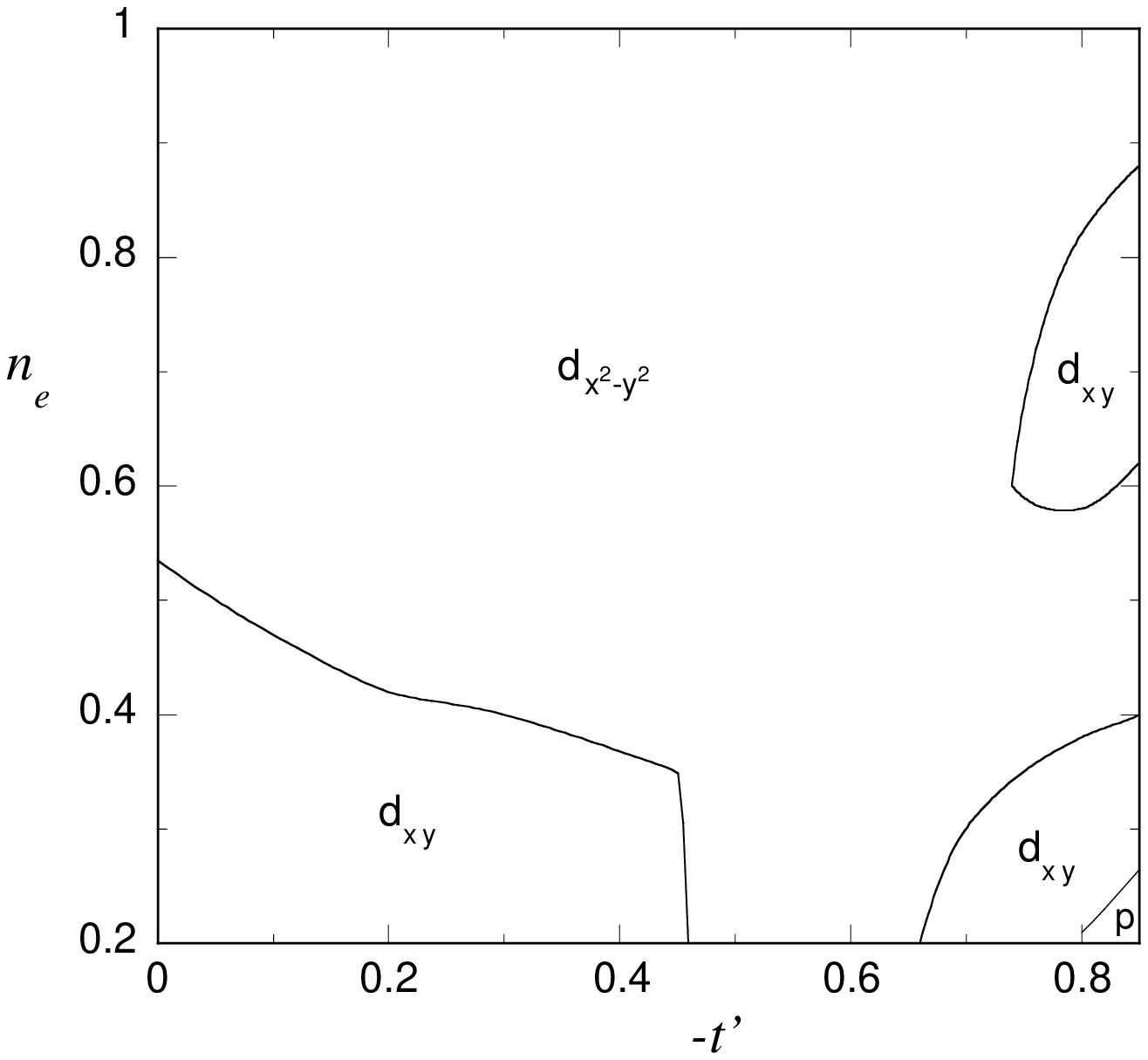}
\caption{
Phase diagram for the square lattice with anisotropic $t'<0$ (lattice of organic 
conductors).
}
\end{center}
\label{phasor1}
\end{figure}

\begin{figure}
\begin{center}
\includegraphics[width=\columnwidth]{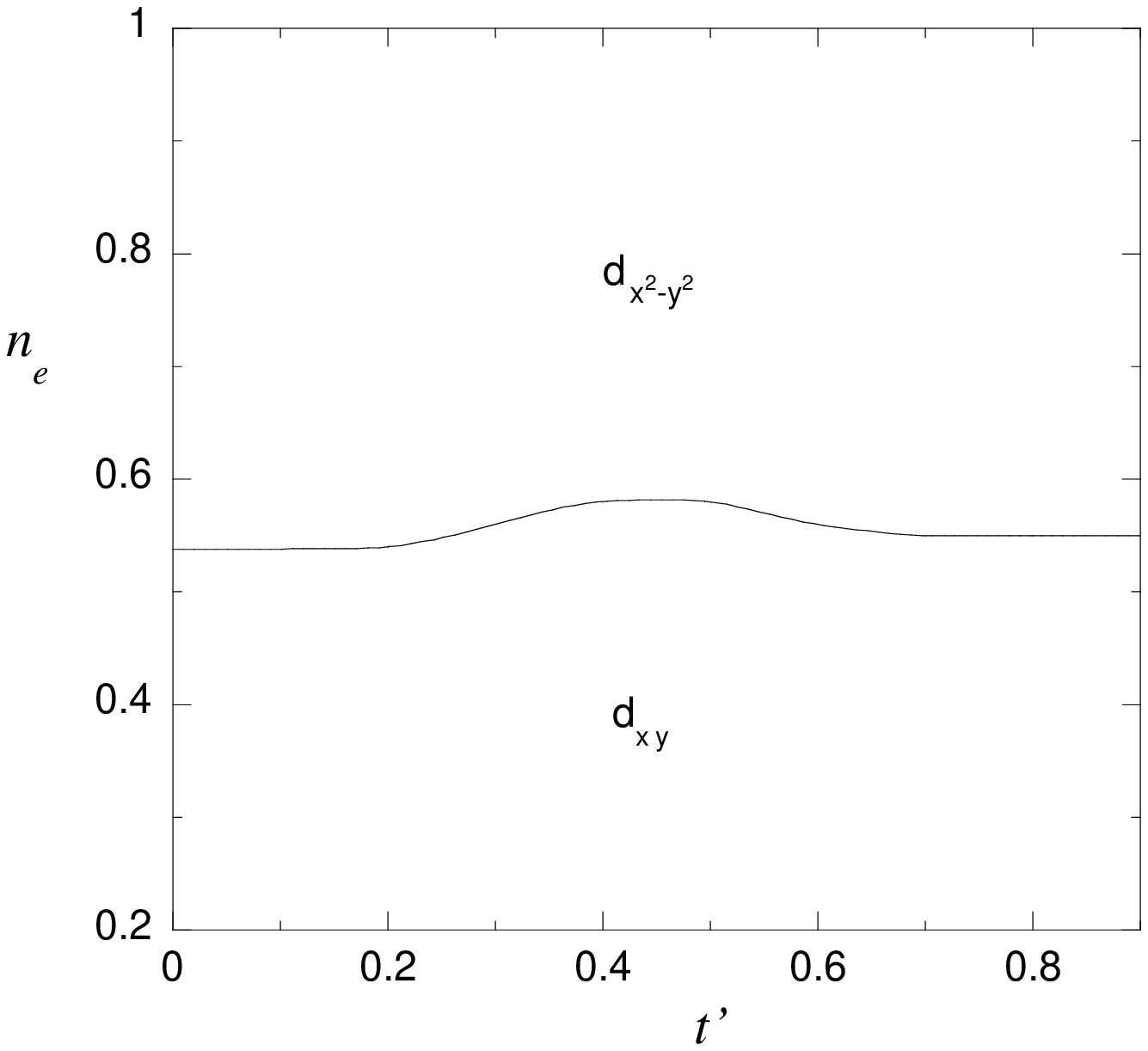}
\caption{
Phase diagram for the square lattice with anisotropic $t'>0$ (lattice of organic 
conductors).
}
\end{center}
\label{phasor2}
\end{figure}

\subsection{Simple square lattice}
Let us investigate the phase diagram for the square lattice (Fig.\ref{latt-sq}).
The basis functions $\{e^{in\theta}\}$ ($n=0,\pm 1,\pm 2, \dots$) are
classified into irreducible representations according to the symmetry
group. 
The eigenfunctions are specified by one of irreducible representations
of the square lattice (see Table 1).
It is convenient to use real basis functions cos$(n\theta)$ and sin$(n\theta)$
for this purpose.
The gap function in each representation is\cite{kon01}
\begin{equation}
z(\theta)= \sum_{\ell=1}z_{4\ell}{\rm cos}(4\ell\theta)~~~~A_{1},
\end{equation}
\begin{equation}
z(\theta)= \sum_{\ell=1}z_{4\ell}{\rm sin}(4\ell\theta)~~~~A_{2},
\end{equation}
\begin{equation}
z(\theta)= \sum_{\ell=1}z_{4\ell-2}{\rm cos}(4\ell-2)\theta~~~B_{1},
\end{equation}
\begin{equation}
z(\theta)= \sum_{\ell=1}z_{4\ell-2}{\rm sin}(4\ell-2)\theta~~~B_{2}.
\end{equation}
\begin{eqnarray}
z(\theta)&=& \sum_{\ell=1}z_{2\ell-1}{\rm cos}(2\ell-1)\theta~~~E.\nonumber\\
&&~~~~~~~{\rm or}~~{\rm sin}(2\ell-1)\theta
\end{eqnarray}
In Ref.\cite{kon01} the representations A$_1$$\sim$B$_2$ were investigated.
Here the E symmetry for triplet pairing is also examined.
The eigenequation is solved for the above shown basis functions in
the space of each irreducible representation.
The eigenvalue $x$ for $t'=0$ is shown in Fig.\ref{xne} as a function of the electron
density $n_e$.
For $n_e>0.6$ the paired state with $d_{x^2-y^2}$ symmetry is most stable
for $t'=0$.
Since the exponent $x$ sensitively depends on the van Hove singularity,
$x$ is an increasing function of $n_e$ near half filling for $t'=0$.

The exponent $x$ for $t'=-0.1$, $-0.2$ and $-0.3$ is shown in Figs.\ref{xne01}, 
\ref{xne02}
and \ref{xne03}, respectively.
The exponents for small electron filling are not shown here because the high
numerical accuracy is required for exponentially small exponents.
As is shown in the figures, the $d$-wave state is most stable near half-filled
case for $t'$ in the range of $0\leq t'\leq 0.4$.
The position of the van Hove singularity depends on $t'$, and the
peak of $x$ shifts as $-t'>0$ increases (Fig.\ref{xnet'}).
$x$ has a sharp peak showing a logarithmic increase due to the van Hove singularity:
\begin{equation}
x\sim -{\rm log}|\mu-\mu_{vH}|,
\end{equation}
where $\mu_{vH}$ is the chemical potential corresponding to the van Hove
singularity.
The figure suggests higher $T_c$ for small $-t'$.  The antiferromagnetism,
however, may compete and suppress superconductivity near half filling.
Hence we must have a bell-shape critical temperature as a function of
the electron filling.

It was pointed out from the electronic states calculations that the Fermi
surface is much deformed for Tl$_2$Ba$_2$CuO$_6$,\cite{sin92} and 
HgBa$_2$CuO$_4$\cite{sin93} for which the band parameter values must be
assigned as
$t'\sim -0.4$ and $t''\sim 0.1$ (third-neighbor transfer).
Bi$_2$Sr$_2$CaCu$_2$O$_{8+\delta}$ (Bi2212) also has deformed Fermi surface
so that $t'\sim -0.3$ and $t''\sim 0.2$\cite{toh00}.
For these values the optimum doping rate must be larger than that for
La$_{1-x}$Sr$_x$CuO$_4$ (LSCO) for which $t'\sim -0.1$ and $t''\sim 0$.
Experiments, however, indicated that the optimum doping rate is almost the
same for Bi2212 and LSCO\cite{har96}. 
This may be a flaw of the weak coupling formulation, which, however, may not be
completely remedied by the strong coupling treatment since the van Hove
singularity still has a large effect on the critical temperature.
This suggests that we must reexamine the structure of the Fermi surface
of high temperature cuprates.
In particular, the band parameters for Bi2212 will be modified if we
take into account the double layer structure\cite{mce03,fen01}.
The band structure reported by recent studies\cite{fen01,hus03} is well fitted
using smaller $t'$ such as\cite{yama}
\begin{equation}
t'\sim -0.2.
\end{equation}

The phase diagram in the $n_e$-$t'$ plane is shown in Fig.\ref{phase1} for
$t'\leq 0$ and in Fig. \ref{phase2} for $t'\geq 0$.
For $n_e\sim 0.5$ and $-t'\sim 0.4$, there is a possibility that the $p$-wave
superconductivity is realized.  For example, the ruthenate superconducting
material Sr$_2$RuO$_4$\cite{mae94} is sometimes modeled by the one-band Hubbard model
for the $\gamma$ orbital
with $t'\sim -0.4$ and $n_e\simeq 0.67$ after the electron-hole transformation.
The state of these parameters just corresponds to the point within the singlet
region near the boundary
to $p$-wave regions in Fig. \ref{phase1}.
In order to obtain the stable $p$-wave pairing for the parameters corresponding to
Sr$_2$RuO$_4$, we may
need to consider the multi-band structure including $\alpha$ and $\beta$
orbitals\cite{koi03}.
For $t'>0$ we have a large $d$-wave region.

If $t'$ is large and negative, i.e. if $-t'>0.5$, we have the case with two
Fermi surfaces; one is a large Fermi surface (FS1) and 
the other is a small Fermi surface (FS2) inside of the larger one.
In this case we must examine the coupled equation of two gap functions $z^1_k$ and $z^2_k$ 
corresponding to two Fermi surfaces:
\begin{equation}
\frac{2}{N}\sum_{{\bf k}':FS1}\chi^{11}({\bf k}+{\bf k'})z^1_{{\bf k}'}
\delta(\xi_{{\bf k}'})
+\frac{2}{N}\sum_{{\bf k}':FS2}\chi^{12}({\bf k}+{\bf k'})z^2_{{\bf k}'}
= -xz^1_{{\bf k}},
\end{equation}
\begin{equation}
\frac{2}{N}\sum_{{\bf k}':FS1}\chi^{21}({\bf k}+{\bf k'})z^1_{{\bf k}'}
\delta(\xi_{{\bf k}'})
+\frac{2}{N}\sum_{{\bf k}':FS2}\chi^{22}({\bf k}+{\bf k'})z^2_{{\bf k}'}
= -xz^2_{{\bf k}},
\end{equation}
where the symbol $\sum_{k':FSi}$ indicates the summation over the Fermi
surface FS$i$ and $\chi^{ij}({\bf k}+{\bf k}')$ is the susceptibility
for ${\bf k}$ on FS$i$ and ${\bf k}'$ on FS$j$. 
The stable pairing symmetry is also obtained using the electron-hole 
transformation for $t'>0$ for which we have almost only one Fermi surface
even in the electron-doped case.

\begin{table}
\caption{Irreducible representations of $C_{4v}$ for the square lattice.
One of basis functions are also shown.
}
\begin{center}
\begin{tabular}{ccccc}
\hline
Rep.    & Symmetry & Bases &  &  \\
\hline
$A_{1}$ &  $s$   &  1              &   &  cos(4$\theta$)  \\
$A_{2}$ & $g$    &  & $xy(x^2-y^2)$ & sin(4$\theta$) \\ 
$B_{1}$ & $d_{x^2-y^2}$  & cos($k_x$)-cos($k_y$) &  $x^2-y^2$ &  cos(2$\theta$)  \\
$B_{2}$ & $d_{xy}$       & sin($k_x$)sin($k_y$)  & $xy$      &  sin(2$\theta$)  \\
$E$     & $p$ & sin($k_x$),sin($k_y$) & $x,y$ & cos($\theta$),sin($\theta$)  \\ 
\hline
\end{tabular}
\end{center}
\end{table}

\subsection{Square lattice with anisotropic t'}
The Hubbard model on the square lattice with anisotropic next-nearest-neighbor
transfer $t'$ (Fig.\ref{latt-org})
has been investigated intensively as a model for organic conductors such as
BDET-TTF(ET) molecules\cite{mck97,mck98,miya04}
The model for organic conductors is well known as the Hubbard model with
anisotropic next-nearest neighbor transfer $t'$ (which is sometimes called
the anisotropic triangular lattice).
The dispersion relation is
\begin{equation}
\xi_k= -2t({\rm cos}k_x+{\rm cos}k_y)
-2t'{\rm cos}(k_x+k_y)-\mu,
\end{equation}
This model has the two-fold rotational symmetry and we classify the
irreducible representation using the $C_{2v}$ point group (Table 2).
The exponent $x$ is in Figs.\ref{xneor01} and \ref{xneor05} as a
function of the electron density $n_e$ for $t'=-0.1$ and $t'=-0.5$,
respectively.
As apparent from the figures, the $d$-wave state is stable over the whole
region, which is consistent with the FLEX prediction\cite{kon04}.
The phase diagram in the $n_e$-$t'$ plane is presented in Fig.\ref{phasor1}
for $t'<0$ and in Fig.\ref{phasor2} for $t'>0$.
For this model we conclude that the $d$-wave pairing is stable over
the whole range of parameters.

\begin{table}
\begin{center}
\caption{Irreducible representations of $C_{2v}$ for the square lattice
with anisotropic next-nearest-neighbor transfer.
}
\begin{tabular}{cccc}
\hline
Representation & Symmetry & Bases &   \\
\hline
$A_{1}$ & $d_{x^2-y^2}$  &  $x^2$, $y^2$  &  cos(2$\theta$)  \\
$A_{2}$ & $d_{xy}$       &  $xy$          &  sin(2$\theta$)  \\
$B_{1}$ & $p_{x}$        &  $x$           &  cos($\theta$)  \\
$B_{2}$ & $p_{y}$        &  $y$           &  sin($\theta$)  \\
\hline
\end{tabular}
\end{center}
\end{table}

\begin{figure}
\begin{center}
\includegraphics[width=\columnwidth]{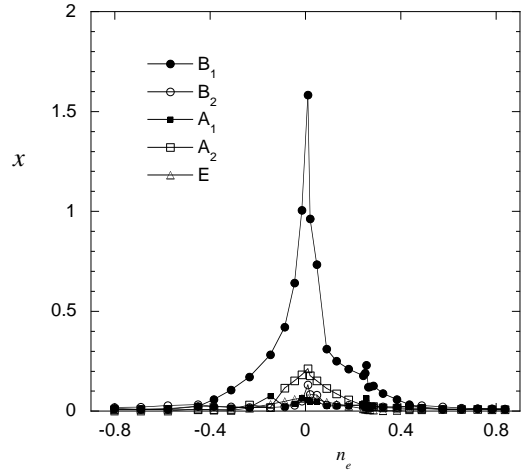}
\caption{
$x$ as a function of the carrier density $n_e$ for the square lattice $d$-$p$ model: 
$n_e>0$ for hole doping
and $n_e<0$ for electron doping.
}
\end{center}
\label{xnedp}
\end{figure}

\begin{figure}
\begin{center}
\includegraphics[width=\columnwidth]{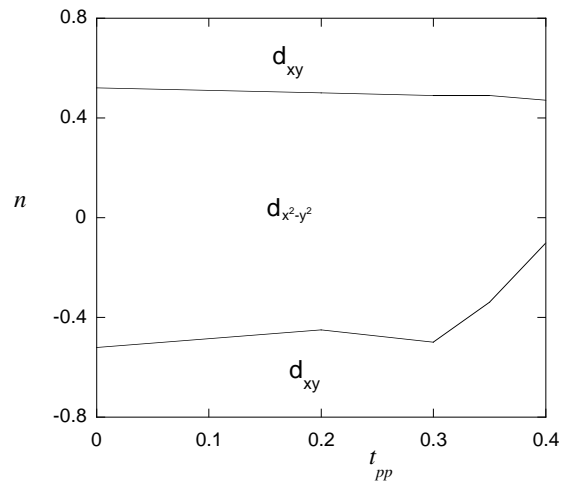}
\caption{
Phase diagram for the three-band $d$-$p$ model in the plane of the carrier
number $n$ and $t_{pp}$ in the range of $0\leq t_{pp}\leq 0.4$. 
We set $\epsilon_p-\epsilon_d=2$
and $t_{dp}=1$.  $n=0$ indicates the half filling, and
the positive and negative $n$ are for hole doping and electron doping,
respectively. 
}
\end{center}
\label{ntpp}
\end{figure}

\subsection{Three-band $d$-$p$ model}

The formulation is also applied to the three-band model for the CuO$_2$
plane\cite{koi01}.
We are interested in the relation between the single-band Hubbard model
and the three-band $d$-$p$ model.
The pairing symmetry in the electron-doped cuprates is still 
controversial between the $d$-wave and $s$-wave order 
parameter\cite{yana01,sat01,che02}.
The Hamiltonian is
\begin{eqnarray}
H_{dp}&=& \epsilon_d\sum_{i\sigma}d_{i\sigma}^{\dag}d_{i\sigma}
+ \epsilon_p\sum_{i\sigma}(p_{i+\hat{x}/2\sigma}^{\dag}p_{i+\hat{x}/2\sigma}
\nonumber\\
&+& p_{i+\hat{y}/2\sigma}^{\dag}p_{i+\hat{y}/2\sigma})
\nonumber\\
&+& t_{dp}\sum_{i\sigma}[d_{i\sigma}^{\dag}(p_{i+\hat{x}/2\sigma}
+p_{i+\hat{y}/2\sigma}-p_{i-\hat{x}/2\sigma}-p_{i-\hat{y}/2\sigma}\nonumber\\
&+& {\rm h.c.}]+t_{pp}\sum_{i\sigma}[p_{i+\hat{y}/2\sigma}^{\dag}p_{i+\hat{x}/2\sigma}
-p_{i+\hat{y}/2\sigma}^{\dag}p_{i-\hat{x}/2\sigma}\nonumber\\
&-&p_{i-\hat{y}/2\sigma}^{\dag}p_{i+\hat{x}/2\sigma}
+p_{i-\hat{y}/2\sigma}^{\dag}p_{i-\hat{x}/2\sigma}+{\rm h.c.}]\nonumber\\
&+& U_d\sum_i d_{i\uparrow}^{\dag}d_{i\uparrow}d_{i\downarrow}^{\dag}
d_{i\downarrow}.
\end{eqnarray}
In this subsection the energy is measured in units of $t_{dp}$.
The energy levels of the non-interacting Hamiltonian is written as
in a concise form\cite{koi01}:
\begin{equation}
\epsilon_{{\bf k}}^{\alpha}= \frac{2}{\sqrt{3}}t_{{\bf k}}{\rm cos}
\left(\frac{\phi_{{\bf k}}+2\pi\alpha}{3}\right)+\frac{\epsilon_d-\epsilon_p}{3}.
\end{equation}
for $\alpha=0$, 1 and 2, where
\begin{equation}
t_{{\bf k}}= \sqrt{|\eta_{{\bf k}}^x|^2+|\eta_{{\bf k}}^y|^2
+(\eta_{{\bf k}}^p)^2+(\epsilon_d-\epsilon_p)^2/3},
\end{equation}
\begin{equation}
\phi_{{\bf k}}= \frac{\pi}{2}+sign(s_{{\bf k}})
\left(\frac{\pi}{2}-{\rm arctan}\sqrt{|1-4t_{{\bf k}}^6/(27s_{{\bf k}}^2)|}
\right),
\end{equation}
\begin{equation}
s_{{\bf k}}=(\epsilon_d-\epsilon_p)\left( \frac{(\epsilon_d-\epsilon_p)^2}{27}
-\frac{t_{{\bf k}}^2}{3}+(\eta_{{\bf k}}^p)^2\right)
+\eta_{{\bf k}}^p(\eta_{{\bf k}}^x\eta_{{\bf k}}^{y*}
+\eta_{{\bf k}}^{x*}\eta_{{\bf k}}^y),
\end{equation}
where $\eta_{{\bf k}}^x=2it_{dp}{\rm sin}(k_x/2)$, 
$\eta_{{\bf k}}^y=2it_{dp}{\rm sin}(k_y/2)$, and 
$\eta_{{\bf k}}^p=-4t_{pp}{\rm sin}(k_x/2){\rm sin}(k_y/2)$.
$\epsilon_{{\bf k}}^{\alpha}$ for $\alpha=0$,1,2 is the dispersion relation
of the upper, lower and middle band, respectively.
We examine the doped case within the hole picture where the lowest band is occupied
up to the Fermi energy $\mu$.
The effective interaction is
\begin{equation}
V_{{\bf k}{\bf k}'}= \frac{U_d}{N}+\frac{U_d^2}{N}\chi^{dd}({\bf k}+{\bf k}'),
\end{equation}
where
\begin{equation}
\chi^{dd}({\bf q})= \frac{1}{N}\sum_{{\bf p}}\sum_{\alpha\beta}
w_{{\bf q}+{\bf p}}^{\alpha}\frac{f_{{\bf q}+{\bf p}}^{\alpha}-f_{{\bf p}}^{\beta}}
{\epsilon_{{\bf p}}^{\beta}-\epsilon_{{\bf q}+{\bf p}}^{\alpha}}
w_{{\bf p}}^{\beta}.
\end{equation}
Here $f_{{\bf k}}^{\alpha}$ is the Fermi distribution function,
\begin{equation}
f_{{\bf k}}^{\alpha}= \left( e^{\beta(\epsilon_{{\bf k}}^{\alpha}-\mu)}+1
\right)^{-1}.
\end{equation}
The weighting factor of $d$ electrons $w_{{\bf k}}^{\alpha}$ is defined as
\begin{equation}
w_{{\bf k}}^{\alpha}= \frac{(\eta_{{\bf k}}^p-\epsilon_{{\bf k}}^{\alpha})
(\eta_{{\bf k}}^p+\epsilon_{{\bf k}}^{\alpha})}
{(\epsilon_{{\bf k}}^{\beta}-\epsilon_{{\bf k}}^{\alpha})
(\epsilon_{{\bf k}}^{\alpha}-\epsilon_{{\bf k}}^{\gamma})},
\end{equation}
where $\alpha$, $\beta$ and $\gamma$ are different from each other. 
The gap equation is
\begin{equation}
\Delta_k= -\sum_{k'}w_kV_{kk'}^{dd}w_{k'}\Delta_{k'}
\frac{1}{2E_{k'}},
\end{equation}
where $w_k=w_{{\bf k}}^1$ and $E_k=\sqrt{\xi_k^2+\Delta_k^2}$
for the lowest-band dispersion $\xi_k=\epsilon_k^1-\mu$.

$d$-wave pairing is predominant over the whole range in the parameter
space as is shown in Fig.\ref{xnedp}.  
In particular, $d_{x^2-y^2}$-wave pairing is stable
near half-filling.  Although the extended $s$-wave pairing is possible
in the narrow region near half filling in the Gutzwiller variational
Monte Carlo study\cite{yan01}, we have no chance of $s$-wave 
superconductivity within the weak-coupling perturbation theory.
The phase diagram for the $d$-$p$ model is shown in Fig.\ref{ntpp}.

\section{Summary}

We have examined the phase diagram with respect to pairing symmetry
on the basis of the two-dimensional Hubbard model.
The weak coupling formulation is convenient to investigate the phase
diagram in detail.  The results are almost consistent with the
strong-coupling perturbation theory.
We summarize the results as follows.\\
(1) The $d$-wave pairing is stable near half filling for the square lattice
and the anisotropic square lattice.\\
(2) The gap function has a maximum at the van Hove singularity.
As the second neighbor transfer $t'$ increases, the energy of the van Hove
singularity decreases.  For large $t'= -0.3\sim -0.4$, the optimal doping
is more than 25 percent doping, i.e. $n_e<0.75$. 
For small third neighbor transfer $t''$ the situation remains the same.
The large $-t'$ is assigned to several high-temperature cuprates to fit
the angle resolved photoemssion spectroscopy (ARPES)
data or the Fermi surface obtained by the band structure calculations.
Most of them, however, have optimum critical temperature in the range of
$0.8<n_e<0.85$.  Thus the weak coupling analysis suggests that we must consider 
other electronic or lattice
interactions, or reexamine the band parameters $t'$ and $t''$.
Recent ARPES studies have reported the band structure which is well fitted
using rather smaller $t'$ such as $t'\sim -0.2$ by our analysis.
\\
(3) The predictions of the weak-coupling theory are almost consistent
with the variational Monte Carlo method.
An effective interaction to induce superconductivity is possibly 
the simple $\chi({\bf q})$ with renormalization in the Gutzwiller 
variational theory.\\
(4) For the $d$-$p$ model, the $d$-wave pairing is predominant in the
wide range and the phase diagram is almost symmetric between electron
and hole dopings.
Although the pairing symmetry in the electron-doped cuprates is 
controversial, only the $d$-wave pairing is possible near half-filling
in the weak-coupling formulation.  

This work was supported by Grants-in-Aid for Scientific Research from
the Ministry of Education, Culture, Sports, Science and Technology of 
Japan.
A part of numerical calculations was performed at the facilities of the
Supercomputer Center of Institute for Solid State Physics, University of Tokyo.

The author expresses his sincere thanks to J. Kondo, K. Yamaji and S. Koikegami
for fruitful discussions.

\appendix
\section{Higher-order corrections}

\begin{figure}
\begin{center}
\includegraphics[width=\columnwidth]{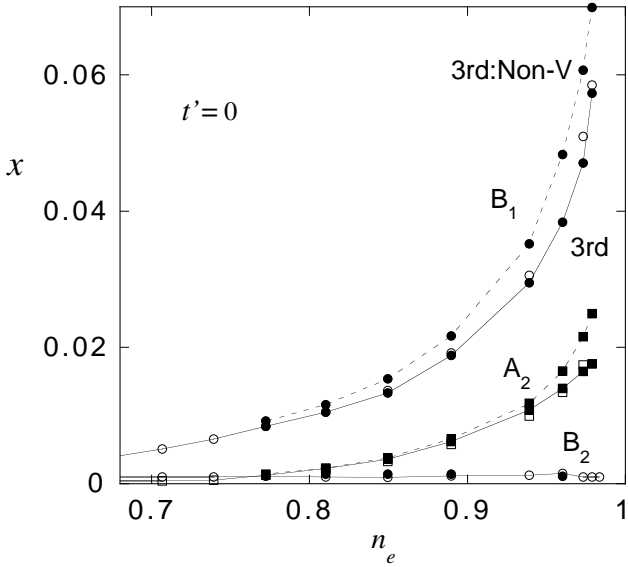}
\caption{
The exponent $x$ for the second-order (open symbols) and third-order
(solid symbols)
perturbation in $U$.
We set $U/t=0.1$.  The symbol Non-V indicates the results obtained without
vertex corrections.
}
\end{center}
\label{x3rd}
\end{figure}

In the Appendix we examine higher-order corrections to $x$.
If the third-order terms have an effect to reduce the exponent $x$,
the results obtained using the second-order perturbation have a 
possibility to become unstable as $U$ increases.
It is not an easy task to derive an effective Hamiltonian up to the
third order of the interaction using the canonical transformation.
The gap equation up to the third order of $U$ has been obtained
using the perturbative expansion for the Hubbard model\cite{juj99,nom01}.
The Green's functions satisfy the Dyson equations:
\begin{eqnarray}
G({\bf k},i\epsilon_n)&=& G_0({\bf k},i\epsilon_n)
+G_0({\bf k},i\epsilon_n)\Sigma_n({\bf k},i\epsilon_n)G({\bf k},i\epsilon_n)
\nonumber\\
&+&G_0({\bf k},i\epsilon_n)\Sigma_a({\bf k},i\epsilon_n)F^*({\bf k},i\epsilon_n),
\end{eqnarray}
\begin{eqnarray}
F({\bf k},i\epsilon_n)&=&
G_0({\bf k},i\epsilon_n)\Sigma_n({\bf k},i\epsilon_n)F({\bf k},i\epsilon_n)
\nonumber\\
&-&G_0({\bf k},i\epsilon_n)\Sigma_a({\bf k},i\epsilon_n)G(-{\bf k},-i\epsilon_n),
\end{eqnarray}
where $\epsilon_n=(2n+1)\pi k_BT$ is the Matsubara frequency, and $\Sigma_n$
($\Sigma_a$) is the normal (anomalous) self-energy.
$G_0$ is the free-electron Green's function:
$G_0({\bf k},i\epsilon_n)=(i\epsilon_n-\xi_k)^{-1}$.
Since we are interested in the third-order contributions, $\Sigma_n$ (of the
order of $U^2$) is neglected as follows:
\begin{equation}
G({\bf k},i\epsilon_n)= -\frac{i\epsilon_n+\xi_k}{\epsilon_n^2+\xi_k^2
+|\Sigma_a({\bf k},i\epsilon_n)|^2},
\end{equation}
\begin{equation}
F({\bf k},i\epsilon_n)= -\frac{\Sigma_a({\bf k},i\epsilon_n)}
{\epsilon_n^2+\xi_k^2+|\Sigma_a({\bf k},i\epsilon_n)|^2}.
\end{equation}
The equation for the anomalous self-energy is
\begin{eqnarray}
&&\Sigma_a({\bf k},i\epsilon_n)= \frac{1}{\beta N}\sum_{k',\epsilon_{n'}}
[U+U^2\chi_0({\bf k}+{\bf k}',i\epsilon_n+i\epsilon_{n'})\nonumber\\
&+&2U^3\chi_0({\bf k}+{\bf k}',i\epsilon_n+i\epsilon_{n'})^2]F({\bf k}',i\epsilon_{n'})
\nonumber\\
&+&U^3\frac{1}{\beta^2N^2}\sum_{k',\epsilon_{n'},p,\epsilon_{\ell}}
G_0({\bf k}',i\epsilon_{n'})[\chi_0({\bf k}+{\bf k}',i\epsilon_n+i\epsilon_{n'})
\nonumber\\
&-&\phi_0({\bf k}+{\bf k}',i\epsilon_n+i\epsilon_{n'})]
G_0({\bf k}+{\bf k}'+{\bf p},i\epsilon_n+i\epsilon_{n'}+i\epsilon_{\ell})
\nonumber\\
&\times& F({\bf p},i\epsilon_{\ell})\nonumber\\
&+&U^3\frac{1}{\beta^2N^2}\sum_{k',\epsilon_{n'},p,\epsilon_{\ell}}
G_0({\bf k}',i\epsilon_{n'})[\chi_0(-{\bf k}+{\bf k}',-i\epsilon_n+i\epsilon_{n'})
\nonumber\\ 
&-&\phi_0(-{\bf k}+{\bf k}',-i\epsilon_n+i\epsilon_{n'})]\nonumber\\
&\times& G_0(-{\bf k}+{\bf k}'-{\bf p},-i\epsilon_n+i\epsilon_{n'}-i\epsilon_{\ell})
F({\bf p},i\epsilon_{\ell}),
\end{eqnarray}
for $\beta=1/k_BT$.
The second and third terms originate from the vertex corrections.
$\chi_0({\bf q},i\omega_m)$ and $\phi_0({\bf q},i\omega_m)$ are defined as
\begin{equation}
\chi_0({\bf q},i\omega_m)= -\frac{1}{N}\sum_{k}
\frac{f(\xi_k)-f(\xi_{k+q})}{i\omega_m+\xi_{k}-\xi_{k+q}},
\end{equation}
\begin{equation}
\phi_0({\bf q},i\omega_m)= -\frac{1}{N}\sum_{k} 
\frac{f(\xi_k)-f(-\xi_{-k+q})}{i\omega_m-\xi_k-\xi_{-k+q}},
\end{equation}
where $\omega_m=2\pi mk_BT$.
We assume that $\Sigma_a$ is small and that we can neglect the $\epsilon$-dependence
since we consider the small-$U$ limit. 
We set $\Delta_{{\bf k}}=\Sigma_a({\bf k},\epsilon_n=0)$, then the equation
for $\Delta_{{\bf k}}$ is derived.
We show the results in Fig.\ref{x3rd} for $U/t=0.1$ on the square lattice.
The exponent $x$ slightly decreases due to the third-order corrections.
There is an cancellation among the third order terms.
As has been shown in the literature\cite{juj99}, the vertex corrections reduce
the exponent $x$ and $T_c$ compared to that without vertex corrections.
\\
\\

\end{document}